\documentclass[aps,prd,twocolumn,showpacs,superscriptaddress,nofootinbib]{revtex4-1}

\usepackage{amsmath,latexsym}
\usepackage{graphicx}
\usepackage{amssymb}
\usepackage{hyperref}

\begin{document}

\title{Dynamical properties of Bianchi-I spacetimes in $f(R)$ gravity}

\author{Saikat Chakraborty}
\email{snilch@iitk.ac.in}
\affiliation{Department of Physics,   Indian Institute of Technology, Kanpur 208016, India}
\author{Kazuharu Bamba}
\email{bamba@sss.fukushima-u.ac.jp}
\affiliation{Division of Human Support System, Faculty of Symbiotic Systems Science, 
 Fukushima University, Fukushima 960-1296, Japan}
\author{Alberto Saa} 
\email{asaa@ime.unicamp.br}
\affiliation{
Department of Applied Mathematics,  
 University of Campinas,  13083-859 Campinas, SP, Brazil.}

\date{\today}

\draft

\begin{abstract}
We present  a dynamical analysis in terms of new  expansion-normalized  variables for homogeneous and anisotropic Bianchi-I spacetimes
in $f(R)$ gravity in the presence of anisotropic matter. With a suitable choice of the evolution parameter,   the Einstein's equations are reduced to an  autonomous  5-dimensional system of ordinary differential equations for the new variables. Further restrictions lead to considerable simplifications. For instance, we show that for a large class of functions $f(R)$, which includes  several   cases
commonly considered   in the literature, all the fixed points are polynomial roots, and hence they can be determined with good accuracy
and classified for stability. Moreover, typically for these cases,   any   fixed point corresponding to
isotropic solutions in the presence of anisotropic matter will be unstable. The assumption
of a perfect fluid as source and or the vacuum cases   imply some dimensional reductions and even more
simplifications. In particular, we find that the vacuum solutions of $f(R) =   R^{1+\delta}$, with
$\delta$ a constant, are governed by  an effective bi-dimensional phase space which can be  analytically constructed, leading to an
exactly soluble dynamics. Finally, 
we demonstrate that several results already reported   in the literature can be re-obtained  
in a more direct and easy way  by exploring our dynamical
formulation.  
\end{abstract}

\maketitle

\section{Introduction}

The observable universe can be described by the homogeneous and isotropic Friedmann-Lemaitre-Robertson-Walker (FLRW) metric with a high degree of accuracy. Any inhomogeneity and anisotropy observed is of very small scale and can be satisfactorily  described by cosmological perturbation theory, see, for instance,   
\cite{Mukhanov:1990me, Riotto:2002yw}. Cosmological perturbations are believed to be generated due to quantum fluctuations in the very early universe. On the other hand, the observation of the large scale isotropy of the present universe suggests that the dynamics of the early universe   must include an inherent isotropization mechanism. Since the dynamics of the universe in any relativistic   theory of
gravity is essentially a non-linear system of ordinary differential equations, they can be described,
in general, by constructing a set of autonomous first order system together with an energy constraint,
see \cite{WE,ReviewDS}, for instance, for comprehensive  reviews on the subject.
 We can therefore state, in the language of nonlinear dynamics, that the isotropic solution  describing our universe must arise as an attractor in the space of more general anisotropic solutions at some early epoch of the universe. The most successful paradigm about the early universe, namely  the  inflationary  scenarios 
 \cite{Guth:1980zm,Linde:1981mu}, assumes the existence of a brief epoch of rapid, almost exponential expansion at the very early stage of the universe. An eternally inflating universe is usually described with a   de Sitter solution, characterized by a constant value of the Hubble parameter $H$. For   cosmologically realistic inflationary scenarios, however, a quasi de Sitter epoch is used, wherein the Hubble parameter gradually decreases, leading to a smooth  end to the inflation. A very successful model for  an inflationary period was given by Starobinsky\cite{Starobinsky:1980te}, which employed an $R+\alpha R^2$ Lagrangian for gravity with a positive value of $\alpha$. Indeed, quantum corrections to General Relativity (GR) leading to modifications in Einstein-Hilbert Lagrangian of this type may not be something unexpected in the high curvature regime as in the early universe. The Starobinsky's inflationary scenario,
which is now receiving considerable attention   due to its compatibility with the
  Planck 2018 Results \cite{Aghanim:2018eyx,Akrami:2018odb},  
  is based on the existence of pure vacuum de-Sitter solution in $ R^2$ gravity, whereas the linear term in $R$ actually plays the role of making the evolution a quasi de Sitter one and supressing the inflation after about $70$ e-foldings, see  \cite{Bamba:2015uma,Bamba:2012cp} for further references.

Here, we will be concerned  with the standard  $f(R)$ modified theory of gravity, which is
governed by the  
action 
\begin{equation}
\label{action}
S_= \frac{1}{2\kappa}\int d^4x \sqrt{-g} f(R) 
+S_M  ,
\end{equation}
where $\kappa = 8\pi G$,    $c=\hbar =1 $, and $S_M$ stands for the usual matter contributions to the total
action. The $f(R)$ gravity has been intensively studied as an alternative description
to dark energy and the late acceleration of the universe, see \cite{Capozziello:2003tk,Nojiri:2003ft,Carroll:2003wy} for instance. 
For recent comprehensive  reviews, see 
\cite{LR,RMP,Rev1,Capozziello:2010zz,Capozziello:2011et,Cai:2015emx,Rev2}. 
Most of the works on the dynamics of (\ref{action}) typically  assume a homogeneous and isotropic FLRW model to start with, and do  not take metric anisotropy into account, see, for instance, \cite{Carloni:2004kp,Amendola:2006we,Carloni:2007br,Carloni:2015jla,D1,D2,D3,D4}. Nevertheless, the
dynamics of metric shear for the case of $R^n$ gravity in vacuum and in presence of an isotropic fluid have previously been studied using the $(1+3)$ covariant formalism \cite{Leach:2006br,Goheer:2007wu}, but
the case involving anistropic matter was still lacking in the literature.

In this work, we have extended the so-called expansion-normalized variables \cite{WE}
to write down the dynamical equations of (\ref{action}), for  a homogeneous and anisotropic Bianchi-I metric in the presence of an   anisotropic fluid, as a  5-dimensional system of ordinary differential equations.
 We will show that some further assumptions may lead
to considerable simplifications in the equations, and for several examples we end up with analytically soluble systems. For the sake of illustration, we consider explicitly the case of $f(R) = R^{1+\delta}$.
First, we show that   the formulation of \cite{Leach:2006br,Goheer:2007wu} is
recovered in the isotropic matter limit. Then, we 
  rederive, in a simpler and more direct way, several known results as the existence
of vacuum Kasner-like solutions for  $-\frac{1}{2}\leq \delta \leq\frac{1}{4}$ \cite{Barrow:2005dn,Clifton:2006kc} and some uniqueness and stability properties of the Starobinsky's isotropic inflationary scenario in $R^2$ gravity \cite{BarrowR^2,Maeda:1987xf,Barrow:2006xb}. We also obtain several explicit new results, as the complete dynamical
characterization of    vacuum solutions for the case  $f(R) = R^{1+\delta}$, and  the
instability of isotropic solutions in the presence of anisotropic matter for all $f(R)$ with polynomials fixed points. 

The paper is organized as follows. In section \ref{section 2}, we present the dynamical 
equations  for a Bianchi-I cosmology   for (\ref{action}) 
in the presence of  an anisotropic fluid. 
The isotropic fluid limit is discussed, and we also introduce the new expansion-normalized variables
for the system.  Section \ref{section 4}
is devoted for the applications of our dynamical approach in several explicit examples, and the last section is left for
some concluding remarks.

\section{Bianchi-I cosmology in $f(R)$ gravity with anisotropic fluid}
\label{section 2}

We will consider the
  homogeneous and anisotropic Bianchi-I metric, which can be conveniently cast for our purposes
    in the following   form \cite{Pereira:2007yy,Chakraborty:2018thg,ChakrabortyPalSaa}
    \begin{equation}
ds^2=-dt^2+a^2(t)\sum_{i=1}^3e^{2\beta_i(t)}(dx^i)^2 ,
\label{bianchi-I}
\end{equation}
where $a(t)$ is the average scale factor and the three functions $\beta_i$, which characterize  the anisotropies, are
such that $\beta_1+\beta_2+\beta_3=0$. In our studies, it will be more convenient to
employ the variables
\begin{equation}
\label{beta+}
\beta_\pm = \beta_1 \pm \beta_2.
\end{equation}
 The total amount of anisotropy in the metric (\ref{bianchi-I}) is given by the quantity
\begin{equation}
\label{sigma}
 \sigma^2=\dot{\beta}_1^2+\dot{\beta}_2^2+\dot{\beta}_3^2 = \frac{3}{2}\dot{\beta}_+^2 + \frac{1}{2}\dot{\beta}_-^2.
\end{equation}
For $\sigma=0$, one can show that  the spatial coordinates $x^i$ can be suitably rescaled to recast the   Bianchi-I metric  
in the standard FLRW form. The Ricci scalar for the metric (\ref{bianchi-I}) reads
 \begin{equation}
 \label{scalar}
 R = 6\dot H + 12H^2 + \sigma^2,
 \end{equation}
 where the average Hubble parameter $H$ is given by the standard expression
 \begin{equation}
 H = \frac{\dot a}{a}.
 \end{equation}
We will assume also the presence of an 
 anisotropic barotropic  fluid with 
   energy momentum tensor parametrized as \cite{EoS}  
\begin{equation}
\label{baro}
T_\mu^\nu = {\rm diag}\left( -\rho, p_1, p_2 , p_3  \right) =
 {\rm diag}\left( -\rho, \omega_1\rho,  \omega_2\rho ,  \omega_3\rho  \right),
\end{equation}
and we define the anisotropic    equation of state  as
\begin{equation}
p_i = (\omega +\mu_i)\rho,
\end{equation}
with $i=1,2,3$, where $\omega$ is the average barotropic parameter and 
$
 \omega_i=\omega+\mu_i,$
with $\mu_1+\mu_2+\mu_3=0$ by construction. As in \cite{ChakrabortyPalSaa}, we will parameterize our fluid by the constants $\omega$ and
$\mu_\pm = \mu_1 \pm \mu_2$.

The dynamics of the Bianchi-I metric (\ref{bianchi-I}) under $f(R)$ gravity action (\ref{action}), in the presence
of and anisotropic barotropic  fluid with energy-momentum tensor (\ref{baro}),  can be described by the following set of equations \cite{ChakrabortyPalSaa},
\begin{eqnarray}
&\displaystyle 3H^2=\frac{\kappa}{f^{\prime}}\left(\rho+\frac{Rf^{\prime}-f}{2\kappa}-\frac{3Hf^{\prime\prime}\dot{R}}{\kappa}\right)+\frac{\sigma^2}{2} &,
\label{bianchi_constraint_1}\\
&\displaystyle  2\dot{H}+3H^2=-\frac{\kappa}{f^{\prime}}\left(\omega\rho+\frac{\dot{R}^2f^{\prime\prime\prime}+\left(2H\dot{R} +\ddot{R}\right)f^{\prime\prime}}{\kappa}\right. \nonumber & \\
 & \displaystyle \quad\quad\quad\quad\quad\quad  -\frac{Rf^{\prime}-f}{2\kappa}\Bigg)-\frac{\sigma^2}{2} ,& 
\label{bianchi_dynamic_1}\\
&  \displaystyle \ddot{\beta}_\pm+\left(3H+\frac{\dot{R}f''}{f'}\right)\dot{\beta}_\pm=\frac{\kappa\rho}{F}\mu_\pm,& 
\label{bianchi_dynamic_2}\\
&\displaystyle \dot{\rho}+ \left( 3H\left(1+\omega\right) +  \boldsymbol{\delta}\cdot  \dot{\boldsymbol{\beta}} \right)\rho =0 , &
\label{bianchi_dynamic_345}
\end{eqnarray}
where  $i=1, 2, 3,$ and
 \begin{equation}
 \label{bolddelta}
\boldsymbol{\delta}\cdot  \dot{\boldsymbol{\beta}} = \mu_1\dot{\beta_1}+\mu_2\dot{\beta_2}
+\mu_3\dot{\beta_3} = \frac{3}{2}\mu_+\dot{\beta}_+  + \frac{1}{2}\mu_-\dot{\beta}_-.
 \end{equation}
 Notice that in the presence of a perfect fluid, we will have  $\mu_+=\mu_-=0$ and the 
 two equations (\ref{bianchi_dynamic_2}) for $\beta_+$ and $\beta_-$
 can be substituted with
 \begin{equation}
 \dot{\sigma}+\left(3H+\frac{\dot{R}f''}{f'}\right){\sigma}=0.
 \end{equation}
 In this case, 
there is no anisotropy in the matter sector and  
  the single variable $\sigma$ is sufficient to describe the total amount of metric anisotropy in the
 system. 
 As we can see, in general, we will have four functions of time $H(t),\,\rho(t),\,\beta_\pm(t)$ governing the dynamics.
The existence of the constraint equation (\ref{bianchi_constraint_1}) implies that only three of them are indeed independent. Without loss of generality, we can choose them to be, for instance, $H(t)$ and $\beta_\pm(t)$. Given some
specific  form of the function $f(R)$, they can be determined by solving equations (\ref{bianchi_dynamic_1}) and (\ref{bianchi_dynamic_2}). The fluid energy density $\rho(t)$ can then be found using the energy constraint (\ref{bianchi_constraint_1}). 

\subsection{The expansion-normalized variables}
\label{section 3}

The traditional expansion-normalized variables were initially introduced for a better dynamical
analysis of the standard FLRW model, see \cite{WE} for instance. Here, we will expand the
variables  already
introduced in  in \cite{Leach:2006br,Goheer:2007wu} to include the case of   the anisotropic 
barotropic fluid (\ref{baro}).    
In this regard, let us   introduce the monotonically increasing variable  
\begin{equation}
N= \epsilon\ln a,
\end{equation}
known as the logarithmic time, 
where $\epsilon$ is defined to be $+1$ for expanding universe and $-1$ for a contracting one. 
Without loss of generality, 
we choose the scale factor at $t=0$ to be $a_0=1$. Therefore, as time progresses in the forward (positive) direction, the logarithmic time $N$ becomes positive and goes towards $+\infty$ in case of both   expanding and contracting universes. One can notice that
\begin{eqnarray}
\dot{N}=\epsilon H,
\end{eqnarray}
so that $\dot{N}$ is effectively always positive, justifying the use of $N$ as the dimensionless
evolution    variable for both expanding and contracting universes. On the other hand, around a bounce or a turnaround point, this argument is not valid though and the expanding and contracting branches must
be considered separately. 

The expansion-normalized  dynamical variables suitable for the equations (\ref{bianchi_constraint_1}) - (\ref{bianchi_dynamic_345}) are the following dimensionless combinations 
\begin{eqnarray}
\label{dynm_variables}
&\displaystyle u_1=\frac{\dot{R}f^{\prime\prime}}{f^{\prime}H},\,u_2=\frac{R}{6H^2},\,u_3=\frac{f}{6f^{\prime}H^2},&\\
&\displaystyle\,u_{4}^+=\frac{\dot{\beta_+}^2}{4H^2},\,u_{4}^-=\frac{\dot{\beta_-}^2}{12H^2},\,u_5=\frac{\kappa\rho}{3f^{\prime}H^2}.& \nonumber
\end{eqnarray}
in terms of which the energy constraint   (\ref{bianchi_constraint_1}) reads simply 
\begin{equation}
g  = 1 + u_1-u_2+u_3-u_{4}^+-u_{4}^--u_5=0,
\label{dynm_const}
\end{equation}
from where we have that one of the expansion-normalized    variables can always be eliminated. Unless
otherwise stated, we will always choose the matter content variable $u_5$ to be expressed in terms of the others
dynamical variables. 
The variable 
\begin{equation}
u_4=u_{4}^++u_{4}^- = \frac{ {\sigma}^2}{6H^2}
\end{equation}
is also relevant for our purposes. It is important to stress that the variables $u_{4}^+$ and $u_{4}^-$ are 
both non-negative   by construction.  
Now, let us
introduce the quantity
\begin{equation}
\label{gamma}
\gamma(R)=\frac{f^{\prime}}{Rf^{\prime\prime}},
\end{equation}
which, or course,     contains     the information about the form of $f(R)$.
Knowing the form of $f(R)$, $\gamma$ can be determined in terms of the dynamical variables $u_2$, $u_3$ by inverting the relation
\begin{equation}
\label{rel}
\frac{u_2}{u_3}=\frac{Rf^{\prime}}{f}.
\end{equation}
We will return to the question of the invertibility of (\ref{rel}) in the last section. 
The 5-dimensional system of autonomous first order differential equations fully equivalent to (\ref{bianchi_dynamic_1}) - (\ref{bianchi_dynamic_345})   is 
\begin{eqnarray}
\epsilon \frac{du_1}{dN} &=&  1+u_2-3u_3-u_4-3\omega u_5 \nonumber \\
&& \quad\quad \quad\quad\quad\quad   -u_1\left(u_1+u_2-u_{4 }\right) ,  
\label{u_1_anifl}\\
\epsilon \frac{du_2}{dN}&=& u_1u_2\gamma\left(\frac{u_2}{u_3}\right)-2u_2\left(u_2-u_{4 }-2\right) ,
\label{u_2_anifl}\\
\epsilon \frac{du_3}{dN}&=& u_1u_2\gamma\left(\frac{u_2}{u_3}\right)-u_3\left(u_1+2u_2-2u_{4 }-4\right) ,
\label{u_3_anifl}\\
\epsilon  \frac{du^+_{4 }}{dN}&=&-2  u_{4}^+\left(1+u_1+ u_2- u_{4  }  \right) + {3} \mu_+ {\sqrt{u_{4}^+}}{u_5} , 
\label{u_4+_anifl}\\
\epsilon  \frac{du^-_{4 }}{dN}&=&-2  u_{4}^-\left(1+u_1+ u_2- u_{4 }  \right)  +  \mu_- {\sqrt{3u_{4}^-}}{u_5},    
\label{u_4-_anifl}\\
\epsilon \frac{du_5}{dN} &=& -u_5\left( 3\omega -1 +u_1 +2u_2 - 2u_{4 } \right.  \\
&& \left.  \quad\quad \quad\quad\quad   +
{3} \mu_+ {\sqrt{u_{4}^+}} +  \mu_- {\sqrt{3u_{4}^-}}
\right),\nonumber
\label{u_5_anifl}
\end{eqnarray}
 Notice that differentiating (\ref{dynm_const}) with respect to $N$ and using the equations (\ref{u_1_anifl})-(\ref{u_5_anifl}), we have
\begin{equation}
\epsilon \frac{dg}{dN} = -(u_1 + 2u_2 - 2u_{4}^+ - 2u_{4}^- - 1)g,
\end{equation}
showing that the constraint $g=0$ is indeed conserved along the solutions of   our equations and
the system (\ref{u_1_anifl}) - (\ref{u_5_anifl}) is effectively 5-dimensional. 

The case of $f(R) = R^{1+\delta}$, with $\delta \ne 0$, will be particularly important in our next examples. For this choice of  $f(R)$, one has simply 
\begin{equation}
\label{gamma1}
\gamma = \delta^{-1},
\end{equation}
and the equations (\ref{u_2_anifl}) and  (\ref{u_3_anifl})  can be considerably simplified.
In this case, the right-handed side of the equations (\ref{u_1_anifl}) - (\ref{u_5_anifl}) involves
only second degree polynomials in $u_1$, $u_2$, and $u_3$, and forth degree in  $\sqrt{u_{4}^-}$ 
and $\sqrt{u_{4}^+}$. Hence, the task of finding the fixed points of our system  
reduce to finding polynomial roots, which may be performed in general with good accuracy. 
Notice that there are other relevant choices for $f(R)$ leading to polynomial fixed points. Besides of the 
trivial extension $f(R)=\alpha R^{1+\delta} + \Lambda$, with $\alpha$ and $\Lambda$ constants,
for which (\ref{gamma1}) also holds. We have also the case $f(R)=\alpha\ln R + \Lambda$, 
which corresponds to $\delta \to -1$ in (\ref{gamma1}).
For the so-called exponential gravity \cite{exp0,exp1,exp2,exp3}, for which $f(R)=e^{\alpha R}$, we have
\begin{equation}
\label{exp}
\gamma = \frac{u_3}{u_2},
\end{equation} 
and the polynomial nature of the fixed points if of course maintained. The same occurs to the 
well known case \cite{Carroll:2003wy} $f(R) = R + \frac{\alpha}{R}$, for which
\begin{equation}
\gamma = \frac{u_2}{u_3-u_2}.
\end{equation}
This case belongs, in fact, to the more general class of functions $f(R) = R^a +  {\alpha}{R^b}$,
with $a\ne b$ constants, 
for which we have
\begin{equation}
\gamma = \frac{u_2}{(b+a-1)u_2 - abu_3}.
\end{equation}
Notice that, as in the exponential case, the function $\gamma$ does not depend on the parameter $\alpha$. This, of course, does not mean that the dynamics in insensitive to the value of $\alpha$, since the
expansion-normalized variables (\ref{dynm_variables})  depend explicitly on $\alpha$.
The case $a=1$ and $b=2$ is the original Starobinsky inflationary scenario \cite{Starobinsky:1980te}, and for the vacuum case our approach reduces to that one considered recently in
\cite{Sta2}. 
In the last section, we will discuss in more detail the vast   class of functions $f(R)$  with polynomial fixed points.

\section{Applications} 

For the sake of illustration, we will consider some explicit examples for our approach.  Some new results
will be obtained, and some other  well known results will be rederived in a simpler and more direct way.
We will consider in this section the case of expanding universes ($\epsilon=1$). Contracting
universes ($\epsilon = -1$) 
 correspond to logarithmic time-reversed dynamics.

\label{section 4}
\subsection{$R^{1+\delta}$ vacuum solutions}

Our first example will be the case $f(R) = R^{1+\delta}$, whose main motivations from a cosmological
perspective can be found in \cite{Leach:2006br,Goheer:2007wu,Barrow:2005dn,Clifton:2006kc}, for instance.  The case with $\delta =0$
is obviously pure GR, for which   
 the corresponding system is lower-dimensional, and our approach simply does not apply. The case logarithmic case $f(R)=\ln R$ must be  treated separately. Hence, we will start
considering $\delta\ne 0$ and $\delta \ne -1$. 
 Since we will deal with vacuum solutions, we set $u_5=0$ in the   equations  (\ref{dynm_const}) and (\ref{u_1_anifl})-(\ref{u_5_anifl}). In this case, notice that (\ref{u_4+_anifl})
 and (\ref{u_4-_anifl}) can be combined in only one equation for $u_4$. We can then use (\ref{dynm_const})
 to write $u_3$ as 
\begin{equation}
u_3=u_2-u_1+u_4-1,
\label{dynm_const_vac}
\end{equation}
and we are left with only three dynamical variables $u_1, u_2,$ and $u_4$.
 Now, there is an interesting point to notice \cite{Carloni:2004kp} about the specific  choice   $ f(R)= R^{1+\delta}$, with $\delta \ne -1$,  namely that
\begin{equation}
\label{u2u3}
\frac{u_2}{u_3}=\frac{Rf^{\prime}}{f}= 1+\delta,
\end{equation}
which combined with the constraint (\ref{dynm_const_vac}) implies  
\begin{equation}
\delta u_2=(1+\delta)(u_1-u_4+1),
\label{dynm_const_mon}
\end{equation}
and we are left in fact with a two-dimensional phase space spanned by  the variables $u_1$ and $u_4$. The corresponding 
dynamical equations in this case are
\begin{eqnarray}
  \frac{du_1}{dN}&=& \phi_1(u_1,u_4) \label{dynm_mon_1} \\
&=& - \delta^{-1}(1+2\delta ){ (  u_1  -u_1^*)(u_1-u_4+1)   } , \nonumber  \\
 \label{dynm_mon_4}
  \frac{du_4}{dN}&=& \phi_4(u_1,u_4)  \\ 
&=& - {2\delta^{-1}(1+2\delta) u_4\left(u_1-u_4+1 \right)}  , \nonumber  
\end{eqnarray}
where
\begin{equation}
\label{u_1^*}
u_1^* =   \frac{2( \delta - 1)}{1+2\delta } .
\end{equation}
The phase space $(u_1,u_4)$ associated with the system (\ref{dynm_mon_1}) - (\ref{dynm_mon_4}) has some
interesting features. For instance, it has
an one-dimensional invariant subspace 
 (a continuous line of fixed points) corresponding to the straight line  $u_1-u_4=-1$. However, from (\ref{dynm_const_vac}) we have
that $u_3 = u_2$ on this line, which implies from  (\ref{dynm_variables}) and (\ref{u2u3}) that
$R=0$ on $u_1-u_4=-1$. Besides of this invariant straight line, we have also the isolated fixed  
  $\left(  u_1^* ,0\right)$, for $\delta \ne -\frac{1}{2}$. The case  
   $\delta=-\frac{1}{2}$ will be  also discussed separately. 

The stability of the isolated   fixed point can be inferred from the linearization of (\ref{dynm_mon_1}) - (\ref{dynm_mon_4}).  The Jacobian matrix of (\ref{dynm_mon_1}) - (\ref{dynm_mon_4}) at the   point   $\left(  u_1^* ,0\right)$ reads
\begin{equation}
\left(\frac{\partial (\phi_1,\phi_4)}{\partial (u_1,u_4)}\right) = -  \delta^{-1}(4\delta - 1) 
\left(
\begin{array}{cc}
 1& 0\\
  0& 2
\end{array}
 \right) ,
\end{equation}
from where we have that such fixed point is stable for $\delta > \frac{1}{4}$ or for $\delta < 0$. For 
the stability of the invariant straight line, we can consider the divergence of the vector field
$(\phi_1,\phi_2)$. One has
\begin{equation}
\label{div}
\nabla\cdot\boldsymbol{\phi} = 
\frac{\partial \phi_1}{\partial u_1} + \frac{\partial \phi_4}{\partial u_4}  = 
\delta^{-1}({(1+2\delta)u_4 + 4\delta - 1} )
\end{equation}
on the invariant line. Recalling that $u_4\ge 0$, we have that the invariant line is   entirely
repulsive (positive divergence)   for $\delta > \frac{1}{4}$  or for $\delta \le -\frac{1}{2}$. 
For $-\frac{1}{2} < \delta \le \frac{1}{4}$, we can have some attractive segments, depending on
the value of $u_4$. We will return to the physical  
  interpretation of this $R=0$ invariant line  in a following sub-section. 
The case $\delta = -\frac{1}{2}$ is particularly curious, since the   isolated fixed point is absent
and we have a second one-dimensional invariant line, namely  $u_1=0$, which is also entirely
repulsive. On the other hand, the case $f(R)=\ln R$ cannot be incorporated in the present analysis since
(\ref{u2u3}) is not valid for $\delta\to -1$, and in fact we have a three-dimensional phase space for such case.   
  
\begin{figure}[t]
 \includegraphics[scale=0.6 ]{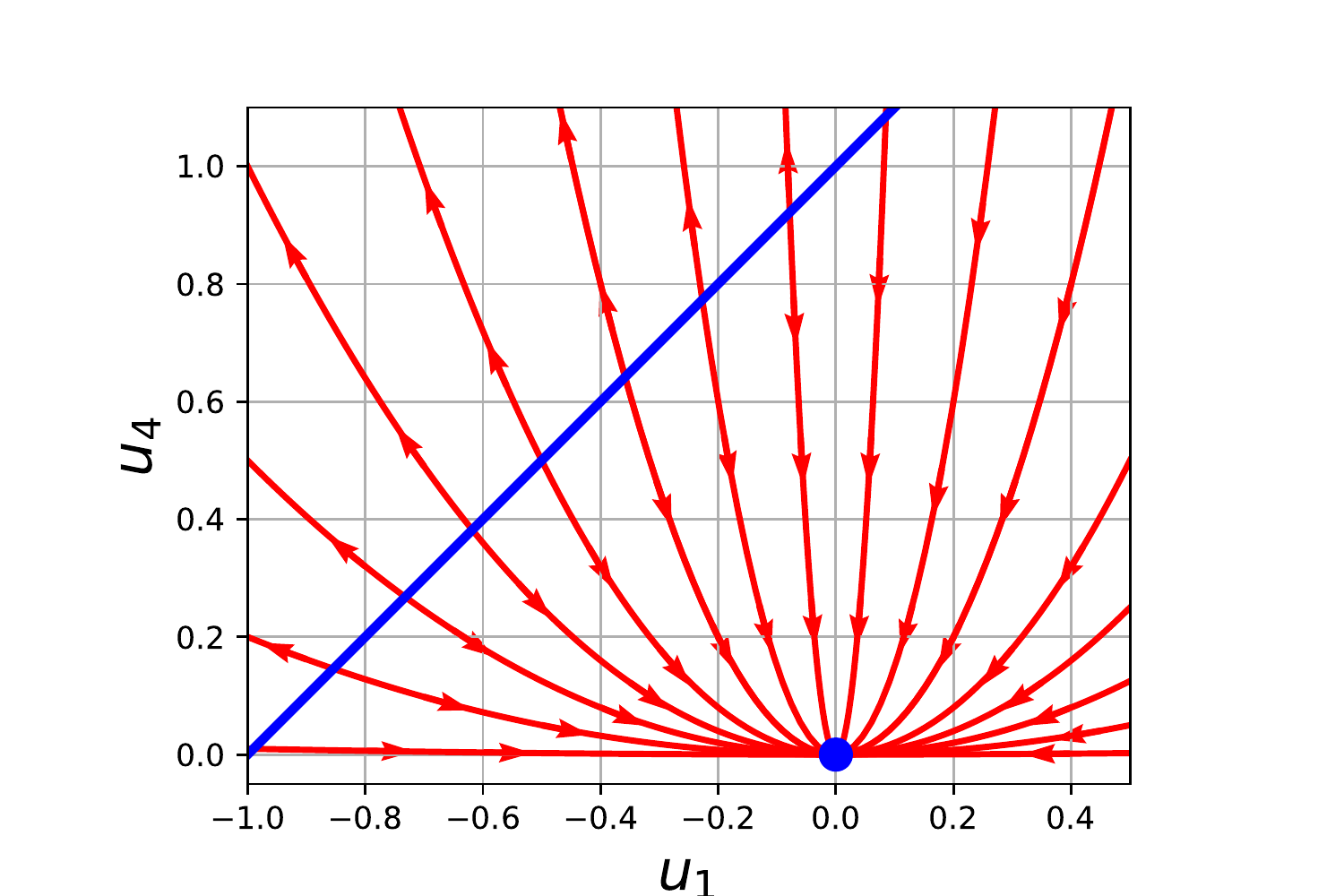}   
\caption{\label{fig1} Phase space for the system (\ref{dynm_mon_1}) - (\ref{dynm_mon_4}), 
for $\delta=1$. The fixed point $(0,0)$ is located in the semiplane below the critical line. The solutions are restricted to parabolas centered in the attractive fixed point.
  The region below the invariant line
corresponds to the attraction basin  of the fixed point. Any solution starting there will
tend asymptotically to the fixed point. All solutions starting in the region above the critical line will diverge to infinity. Notice that the critical line is entirely repulsive. Such phase space is rather generic, it is essentially the same for all theories of the type $f(R)=R^{1+\delta}$ such that the fixed point is attractive and is located below the invariant line. }
\end{figure}   
\begin{figure}[t]
 \includegraphics[scale=0.6 ]{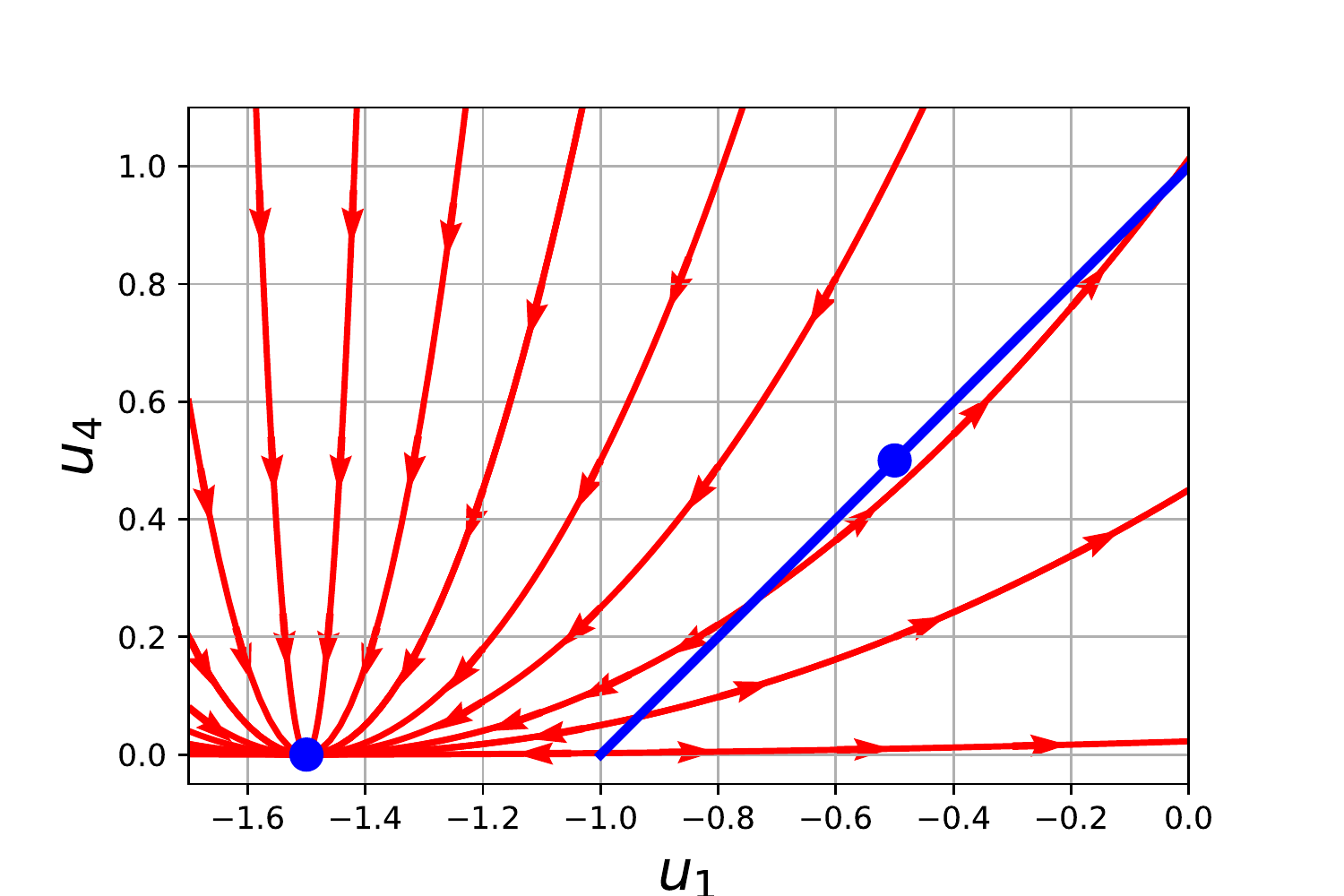}   
\caption{\label{fig2} Phase space for the system (\ref{dynm_mon_1}) - (\ref{dynm_mon_4}), 
for $\delta=\frac{1}{10}$. 
The fixed point $\left(-\frac{3}{2},0\right)$ is now located in the semiplane above
 the critical line.
The solutions are also restricted to parabolas centered in the attractive fixed point. 
However,  the attraction basin of the fixed point is now in the  region above the critical line.
Notice that 
the invariant line in this case has an attractive and a  repulsive segment  located, respectively,
above and below the depicted point $\left(-\frac{1}{2},\frac{1}{2}\right)$.  
The divergence (\ref{div}) always vanishes in   limit points between attractive and repulsive segments like
this one.}
\end{figure}

The solutions of (\ref{dynm_mon_1}) and (\ref{dynm_mon_4}) are curves on the plane $(u_1,u_4)$, and it turns out that such
curves can be determined analytically. Notice that the solutions  are such that
\begin{equation}
\frac{u_4'}{u_1'} = \frac{2   u_4}{ u_1  - u_1^*},
\end{equation}
which can be integrated as
\begin{equation}
\label{para}
u_4 = c\left( u_1  - u_1^* \right)^2,
\end{equation}
with  arbitrary $c$. Thus, the phase space trajectories of all solutions of (\ref{dynm_mon_1}) and (\ref{dynm_mon_4})
are simply parabolas centered in the isolated fixed point, irrespective of   the value of $\delta$, provided the fixed point exists.
Since we known   the trajectories graphs, one can infer
the dynamics direction and, consequently,   
the dynamical properties of the fixed point and the invariant line, directly form the equations 
(\ref{dynm_mon_1}) and (\ref{dynm_mon_4}) as follows. Consider the phase space function $L = u_1-u_4 + 1$. It is clear that $L=0$ is the invariant line. On the other hand, $L=c$ constant is a parallel line located below the invariant line if $c>0$, or above if $c<0$. The invariant line is the boundary between
two semiplanes with reverse dynamics direction, and the dynamical properties of the fixed point
and of the invariant line depend on the relative position between then, see Figs \ref{fig1} 
and \ref{fig2}, which correspond, respectively,
to the cases $\delta =1$ and $\delta=\frac{1}{10}$. The former  is the important case of the Starobinsky's inflationary scenario with $f(R)=R^2$, which we will discuss in more details in the next subsection.   

Notice
that knowing that the solutions are constrained to the parabolas (\ref{para}), the exact solutions of
(\ref{dynm_mon_1}) and (\ref{dynm_mon_4}) boils down to a simple quadrature of a rational function
\begin{equation}
\frac{d\bar u_1}{c\bar u_1^3 - \bar u_1^2 - (u_1^* + 1)\bar u_1} = - \delta^{-1}(1+2\delta) dN ,
\end{equation}
with $u_1 = \bar u_1 + u_1^*$. For the   case $\delta  =-\frac{1}{2}$,
$u_4$ is a constant and (\ref{dynm_mon_1}) also reduces to a simple rational quadrature. 
We have just established that the   vacuum solutions for the  $f(R)=R^{1+\delta}$ 
case, for $\delta\ne - 1 $, are exactly soluble.

Since the stable fixed points of a cosmological model correspond  to the cosmological histories which will dominate
the asymptotic evolution of the system, it worth to look more closely on them.
By using (\ref{dynm_const_vac}) and (\ref{u2u3}), we have that the isolated   fixed points are such
that
\begin{equation}
\label{u3c}
u_3 = \frac{4\delta  -1}{\delta (1+2\delta)},
\end{equation}
with $\delta  > \frac{1}{4}$ or   $\delta  < 0$. From the definition of $u_3$ and (\ref{scalar}), we have
that (\ref{u3c}) implies that 
\begin{equation}
 {\dot H} = \Delta  {H^2},
\end{equation}
where
\begin{equation}
\Delta = \frac{\delta - 1}{\delta ( 1+2\delta )}.
\end{equation}
 It is clear that for
$\delta = 1$, the stable fixed point corresponds to 
de Sitter solution with $a(t) = e^{Ht}$, with constant $H$. (The case $H=0$ corresponds to the flat
 Minkowski spacetime). This is namely the well known Starobinsky's inflationary solution, which we will
 consider in more details in the next subsection. For $\delta \ne 1$, the solutions are
 \begin{equation}
 \label{solH}
 H(t) = \frac{H_0}{1 -   \Delta H_0 (t-t_0)},
 \end{equation}
 where $H(t_0)=H_0$, which interpretation is straightforward. For $\Delta > 0$, which corresponds to
 $-\frac{1}{2} < \delta < 0$ or $\delta > 1$, 
  we have a future finite time big rip singularity, while for $\Delta <0$ ($\delta <-\frac{1}{2}$
or   $0 < \delta < 1$), the Hubble parameter $H$ decreases as $t^{-1}$ for large $t$, {\em i.e.}, the solution
asymptotically tends to a power law expansion.

\subsection{Uniqueness of Starobinsky's inflationary scenario}
\label{uniq}

From the last subsection, we have that the Starobinsky's $R^2$ inflationary scenario is unique among the
  $F(R)=R^{1+\delta}$ theories of gravity, since only for $\delta  =1$ the stable de Sitter fixed point
$(0,0)$ is available, a result indeed   known for a long time, see  
\cite{BarrowR^2,Maeda:1987xf,Barrow:2006xb}, for instance.  We can, however, easily prove  a stronger result for generic $f(R)$ theories. The de Sitter solution $a(t) = e^{Ht}$,
with constant and arbitrary $H$, implies $u_1=u_4=0$, and also
\begin{equation}
\label{RdS}
R = 12H^2,
\end{equation}
which, on the other hand, determine that $u_2=2$ and $u_3=\frac{2f}{Rf'}$ and, hence, the constraint (\ref{dynm_const}) will read
\begin{equation}
\label{RFR}
Rf'(R) = 2f(R).
\end{equation}
Since we assume that  de Sitter solution   exists for arbitrary $H$, we have from (\ref{RdS})
that it should exist for any $R>0$, and hence equation (\ref{RFR}) can be seen as a ordinary differential
equation for $f(R)$, which unique solution is $f(R)=\alpha R^2$, establishing in this way a stronger result: the case $R^2$ is unique among all vacuum $f(R)$ theories with respect to the existence of a de Sitter
solution with arbitrary $H$. The condition (\ref{RFR}) was first obtained by Barrow and Ottewill in \cite{BarrowR^2} by using a more intricate approach, but here we see that it appears from a very simple
analysis of fixed points. We will return to this problem in the last section.

\subsection{Kasner-type solutions}

The third example that we wish to consider is the vacuum Kasner-type solution of the form
\begin{equation}
ds^2=-dt^2+\left(t^{p_1}dx^1\right)^2+\left(t^{ p_2} dx^2\right)^2+\left(t^{ p_3} dx^3\right)^2
\label{kasner}.
\end{equation}
The Ricci scalar, the average Hubble constant, and the total anisotropy $\sigma$ for such metric read, respectively,
\begin{eqnarray}
R &=&\frac{2}{t^2}\left(Q+S-P \right), \quad 
H = \frac{P}{3t},\quad{\rm and}\nonumber \\
\sigma^2 &=& \frac{1}{3t^2}\left(3Q -P^2 \right),
\end{eqnarray}
where
\begin{eqnarray}
Q&=&p_1^2+p_2^2+p_3^2, \\ 
S&=& p_1p_2 + p_1p_3 + p_2p_3, \\
P&=&p_1+p_2+p_3.
\end{eqnarray}
We will consider  here the case $f(R)=R^{1+\delta}$,  exactly in the same line of \cite{Barrow:2005dn, Clifton:2006kc}. In terms of our expansion-normalized variables, 
the Kasner solutions in this case corresponds to following fixed points
\begin{eqnarray}
u_1 &=& -\frac{6\delta }{P} , \\
u_2 &=&   \frac{3(Q+S-P)}{P^2}, \\
u_3 &=& \frac{3(Q+S-P)}{(1+\delta) P^2}, \\ \label{ku4}
u_4 &=&  \frac{3Q -P^2}{2P^2}
\end{eqnarray}
Notice that we already know from the first example the complete phase space for the vacuum $f(R)=R^{1+\delta}$ theory. It has a fixed isolated point and an invariant straight line. The fixed point corresponds to an isotropic solution ($u_4=0$) and imposing the condition $u_1=u_1^*$, we have
\begin{equation}
p_1=p_2=p_3 = \frac{\delta(1+2\delta)}{1-\delta},
\end{equation}
which is a well known FLRW type solution  for $ R^{1+\delta}$ gravity \cite{iso1,iso2,iso3}; see also \cite{Barrow:2005dn} for further discussions. 
All possible anisotropic Kasner solutions must be necessarily on the zero curvature invariant line  where $u_2=u_3=0$, implying that 
\begin{equation}
\label{c1}
Q+S-P=0.
\end{equation}
 The invariant line $u_1 - u_4 + 1 = 0$ in this case reads
\begin{equation}
\label{c2}
12\delta P + 3Q - 3P^2 = 0. 
\end{equation} 
Notice that there is another algebraic relation valid for all $P$, $Q$, and $S$, namely
\begin{equation}
\label{c3}
P^2 -Q - 2S = 0.
\end{equation}
Solving the equations (\ref{c1}), (\ref{c2}), and (\ref{c3}) for $P$, $Q$, and $S$ gives
\begin{eqnarray}
\label{P}
P &=&  2\delta +1 ,\\ \label{Q}
Q&=&  (2\delta  + 1)(1-2\delta),\\
S&=& 2\delta (2\delta  + 1) .
\end{eqnarray}
Since $Q\ge 0$, we have from (\ref{Q}) that the existence of a Kasner solution  requires $-\frac{1}{2}\le \delta \le \frac{1}{2}$, where both limits correspond  to the Minkowski spacetime ($p_1=p_2=p_3=0$). However, there is another more restrictive condition, namely the positiveness
of $u_4$ given by (\ref{ku4}) 
\begin{eqnarray}
3Q-P^2 = 2(2\delta +1)(1-4\delta) \ge 0,
\end{eqnarray}
from where we have   $-\frac{1}{2}\le \delta \le \frac{1}{4}$,
which is exactly   
Barrow and Clifton's result, originally obtained in a more intricate way in \cite{Barrow:2005dn, Clifton:2006kc}. The stability of these solutions is a quite interesting issue. First, notice that 
$
\nabla\cdot\boldsymbol{\phi} = 0
$
at the point
$\left( -\frac{6\delta}{1+2\delta}, \frac{1-4\delta}{1+2\delta} \right)$ of the invariant line, which
means that we need to go further the linear analysis in this case. However, the restriction for the
existence of this Kassner-like solution implies that we are in a situations as depicted in Fig.
\ref{fig2}, with the fixed point above the invariant line. It turns out that the Kassner-like solution
always corresponds to the limit point of the attractive and repulsive segments of the invariant line(!),
implying that both eigenvalues of the Jacobian matrix at this point vanish. However, it is clear
the non-linear instability of this point, since any point above the invariant line is in the
attraction basin of the fixed point.

\subsection{Fixed points with anisotropic matter} 

As an example of application of the full set of our expansion-normalized variables, let us consider
the case of $f(R)$ theories with polynomials fixed points. The simplest case is our example   of a $f(R) = R^{1+\delta} $ theory, but now with an anisotropic barotropic fluid of the type
(\ref{baro}). Taking into account (\ref{gamma1}) and (\ref{u2u3}), the full set of equations in this case will be 
\begin{eqnarray}
 \frac{d u_1}{dN}&=&  1+(\delta -2)u_3-u_4-3\omega u_5 \nonumber \\
&& \quad\quad \quad\quad    -u_1\left(u_1+(1+\delta)u_3-u_{4 }\right) , 
\label{ku_1_anifl}\\
 \frac{d u_3}{dN}&=&  u_3\left(    \delta^{-1}u_1 -2(1+\delta)u_3 +2u_{4 }+4\right) ,
\label{ku_3_anifl}\\
   \frac{d u_{4}^+}{dN} &=&-2  u_{4}^+\left(1+u_1+ (1+\delta)u_3- u_{4  }  \right)  \nonumber \\
&& \quad\quad \quad\quad\quad\quad \quad\quad\quad + {3} \mu_+ {\sqrt{u_{4}^+}}{u_5},
\label{ku_4+_anifl}\\
  \frac{d  u_{4}^-}{dN} &=&-2  u_{4}^-\left(1+u_1+ (1+\delta)u_3- u_{4 }  \right) \nonumber \\
&& \quad\quad \quad\quad\quad\quad \quad\quad \quad+  \mu_- {\sqrt{3u_{4}^-}}{u_5}, 
\label{ku_4-_anifl}
\end{eqnarray}
recalling that $u_4 = u_{4}^++u_{4}^-$, and 
\begin{equation}
u_5 = 1 + u_1-\delta u_3-u_{4 } .
\end{equation}
We are particularly interested in the isotropic fixed points, {\em i.e.}, the solutions with
$u_{4}^+=u_{4}^-=0$. It turns out that there exist four isolated fixed points of this type, namely 
the following values for the pair $(u_1,u_3)$
\begin{eqnarray}
&\displaystyle\left(-1,0 \right), \quad 
\left(1-3\omega,0 \right), \quad
\left(   \frac{ 2(\delta - 1)}{1+2\delta },\frac{4\delta -1}{\delta(1+2\delta)} \right),& \nonumber \\
&\displaystyle \left(-\frac{3\delta(\omega+1)}{1+\delta},\frac{4\delta +1 -3\omega}{2(1+\delta)^2} \right).&
\end{eqnarray}
The relevant question here is whether some of these fixed points are attractive, which would correspond
to asymptotically stable isotropic solutions in the presence of anisotropic matter. We will show that for anisotropic
fluids, all isotropic fixed points are unstable. Nevertheless, for isotropic fluids ($\mu_+=\mu_-=0$), in principle, some of the isotropic fixed points could be indeed stable. 

In order order to prove that the system (\ref{ku_1_anifl}) -  (\ref{ku_4-_anifl}) do not admit any
stable isotropic fixed point in the presence of anisotropic matter, let us assume, without loss of generality, that $\mu_+> 0$, and
consider  (\ref{ku_4+_anifl}) near a generic isotropic fixed point $ (u_1,u_3,u_{4}^+,u_{4}^-)=(u_1^*,u_3^*,0,0)$,  
\begin{equation}
\label{near0}
\frac{du_{4}^+}{dN}  = \sqrt{u_{4}^+}\left( 3\mu_+u_5^* -2\sqrt{u_{4}^+} \left( 1+u_1^* + (1+\delta)u_3^*\right) \right),
\end{equation}
where $ u_5^* = 1 + u_1^*-\delta u_3^*$ is the matter content associated with the fixed point. For the
case of a barotropic anisotropic fluid (\ref{baro}), it is natural to assume $u_5^*> 0$.
Since $\mu_+u_5^*>0$, it is clear that   there is a neighborhood  of $u_{4}^+=0$ where the right-handed side 
 of (\ref{near0}) is positive, implying the repulsiveness of the isotropic fixed points at least
along the positive $u_{4}^+$ direction. For the case
 of an isotropic fluid, since $\mu_+=\mu_-=0$, one can have attractive fixed points according to the sign of the term between parenthesis.  Nevertheless, for anisotropic fluids, no isotropic fixed point can be  stable. 
It is important to stress that, from the structure of the equations (\ref{u_1_anifl}) - (\ref{u_5_anifl}),
we see that the same conclusion will hold for any choice of $f(R)$ with polynomial fixed points, since
we will always have a   repulsive neighborhood  of $u_{4}^+=0$ as we had in (\ref{near0}). 

\subsection{Exponential gravity}

As a last example for our dynamical approach, let us consider the case of exponential gravity $f(R)=e^{\alpha R}$, $\alpha\ne 0$,
which main motivations and implications in cosmology can be found, for instance, in \cite{exp0,exp1,exp2,exp3}.
Taking into account (\ref{exp}), we will have in this case the   5-dimensional system given by the
equations (\ref{u_1_anifl})-(\ref{u_4-_anifl}), but now with the following governing equations for
$u_2$ and $u_3$
\begin{eqnarray}
\frac{du_2}{dN}&=& u_1u_3-2u_2\left(u_2-u_{4 }-2\right) ,
\label{kku_2_anifl}\\
 \frac{d u_3}{dN}&=& -2 u_3\left(  u_2-u_4 - 2\right) ,
\label{kku_3_anifl}
\end{eqnarray}
recalling that $u_4 = u_{4}^++u_{4}^-$. Notice that for the exponential gravity, the phase space variable $u_3$ reads
\begin{equation}
 u_3 = \frac{1}{6\alpha H^2}
\end{equation}
and, hence, it ranges over $(0,\infty)$ and $(-\infty,0)$, respectively, for $\alpha>0$ and $\alpha<0$. The variable $u_4$ is non-negative and all other variables can assume any real value. This is the phase space for the exponential gravity theory.

Let us consider first the vacuum case, for which $u_5=0$, implying that equations (\ref{u_4+_anifl}) and (\ref{u_4-_anifl}) can be combined into a single equation for $u_4$,  and that $u_2$ can be eliminated by using (\ref{dynm_const}), leading finally to the following
3-dimensional system
\begin{eqnarray}
 \frac{d u_1}{dN}&=&   2+u_1  - 2u_3 -2u_4 \nonumber \\
&& \quad\quad \quad    -u_1\left( 1+2u_1+u_3-2u_4\right) , 
\label{kkku_1_anifl}\\
 \frac{d u_3}{dN}&=&  2 u_3\left(  1 -u_1 -u_3 +2u_4\right) ,
\label{kkku_3_anifl}\\
   \frac{d u_{4}}{dN} &=&-2  u_{4}\left(2+2u_1+ u_3- 2u_{4 }  \right)  .    
\label{kkku_4_anifl}
\end{eqnarray}
It is clear from (\ref{kkku_1_anifl}) - (\ref{kkku_4_anifl}) that the boundary plane $u_3=0$ is an invariant
subspace, implying that no solution will ever reach it in finite time. Any solution on this invariant 
subspace can be only reached asymptotically for $N\to\infty$. A fixed-point analysis of our system  reveals the existence of the following
fixed points $(u_1,u_3,u_4)$:
\begin{equation}
(u_4-1,0,u4), \quad (1,0,0), \quad (0,1,0).
\end{equation}
The first solution is exactly the same invariant subspace we have already discussed in our first two  examples, which in the present case is entirely   contained 
in the invariant boundary $u_3=0$. The other two solutions are isolated isotropic fixed points whose cosmological history can be reconstructed from (\ref{scalar}) and, for instance, from the definition of $u_2$ given 
by (\ref{dynm_variables}), which implies
\begin{equation}
\dot H = ( u_2-2)H^2.
\end{equation}
Taking into account (\ref{dynm_const}), we have that
that the invariant line corresponds to a cosmological history like (\ref{solH}), while  
 both isolated fixed points are de Sitter solutions. The respective eigenvalues for these fixed points are
\begin{equation}
(-8,-4,0)\quad {\rm and}\quad \left( -6, -\frac{3+\sqrt{17}}{2}, \frac{ \sqrt{17}-3}{2}\right),
\end{equation}
revealing that the fixed point $(0,1,0)$ is clearly unstable (a saddle point). For the $(1,0,0)$ 
fixed point at the boundary,
the indifferent direction is $\boldsymbol{v}=(3,-4,0)$. However, from a closer inspection of the system (\ref{kkku_1_anifl}) - (\ref{kkku_3_anifl}) 
on the line $(1+3s,-4s,0)$  
\begin{equation}
\boldsymbol{v}\cdot  \frac{d \boldsymbol{u}}{dN} = 
3 \frac{d u_1}{dN} - 4 \frac{d u_3}{dN} = 14s^2,
\end{equation}
we conclude that such fixed point ($s=0$) will be indeed stable for $u_3>0$
($\alpha > 0$) and unstable for $u_3<0$ ($\alpha<0$).
These results are entirely compatible with the spatially flat case  of those ones obtained
in \cite{exp0} for the isotropic case of exponential gravity. 

Let us now consider the case of exponential gravity in the presence of anistropic matter of the
type (\ref{baro}), in the same line we have followed in the preceding subsection. We are also interested
here in the isotropic fixed points $u_4^+ = u_4^-=0$. The dynamical equations have five isolated
isotropic fixed points in this case, namely the values of $(u_1,u_2,u_3)$ given by
\begin{eqnarray}
&(1,2,0), \quad (-3(1+\omega),2,0), &  \\
&  (-1,0,0),\quad (1-3\omega,0,0), \quad
(0,2,1).& \nonumber
\end{eqnarray}
As in the previous example, the relevant question here is whether any of these
fixed points is attractive, which would correspond to
an asymptotically stable isotropic solution  in the presence
of anisotropic matter. It turns out that exactly the same results of the preceding subsection
hold  here, since the structure of the equations for $u_4^+$ and $u_4^-$ are essentially the
same   for  exponential gravity  and for any theory with polynomial fixed points as, for instance, $f(R)=R^{1+\delta}$. All isotropic fixed points for
exponential gravity in the presence of an anisotropic barotropic fluid are unstable, even though
for   isotropic fluids $(\mu_+=\mu_-=0)$ one,  in principle, 
might have some stable isotropic fixed points.

\section{Final Remarks}
\label{section 5}

We have introduced a new set of expansion-normalized variables for  
 homogeneous and anisotropic Bianchi-I spacetimes
in $f(R)$ gravity in the presence of anisotropic matter. In terms of these
new dynamical variables, the full set  of Einstein's equations boils down  
 to a 5-dimensional phase space. As applications of the proposed dynamical approach,
  we have considered explicitly 
the  $f(R)=R^{1+\delta}$ modified theory of gravity, and   shown that its vacuum dynamics is exactly solvable. We have re-obtained, in a easier and more direct way, several well known results for
this particular choice of $f(R)$, as, for instance, Bleyer and Schmidt isotropic solutions \cite{iso1,iso2,iso3} and Barrow and Clifton anisotropic ones \cite{Barrow:2005dn,Clifton:2006kc}. We have also extended a
uniqueness result for Starobisnki inflationary scenario, namely that the case $R^2$ is unique among all vacuum $f(R)$ theories with respect to the existence of a de Sitter
solution with arbitrary $H$, a result obtained previously by Barrow and  Ottewill
by using a more intricate approach  \cite{BarrowR^2}. Finally, we explore our full set
of equations and demonstrate  that, in the presence of anisotropic barotropic 
fluids of the type (\ref{baro}), no isotropic fixed point can be  stable for $f(R)$ theories gravity
with polynomials fixed points. The case of exponential gravity \cite{exp0,exp1,exp2,exp3} was also
explicitly treated. 

There are several possibilities of applications for our dynamical formulation. For instance,  
we could extend the results 
on the existence of de Sitter solution 
of Section \ref{uniq}    for other situations. Goedel and
Einstein universes are natural candidates, since there  already exist  some existence
results in the literature \cite{GE}. We could also consider different geometric situations as,
for instance, the case of Bianchi-IX metrics \cite{IX} or the presence of torsion\cite{tor1,tor2}.
Some of these points are now under investigation. 

As a final remark, let us consider the 
two related issues   left behind in the previous analysis, namely the full set of function $f(R)$ leading
to polynomial fixed points, and the question of the invertibility of (\ref{u2u3}). Firstly, notice that
the condition
\begin{equation}
\label{last}
\gamma = \frac{f'}{Rf''} = c\left(\frac{u_2}{u_3} \right)^{q} = c \left(\frac{Rf'}{f} 
\right)^{q},
\end{equation}
with rational $q$  and $c$ constants, is sufficient to assure that 
 the fixed points of the system (\ref{u_1_anifl}) - (\ref{u_5_anifl}) will be   polynomial roots. The cases
 we have considered, $f(R)=R^{1+\delta}$ and $f(R)=e^{\alpha R}$, correspond, respectively, to the choices $q=0$, $c = \delta^{-1}$ and $q=-1$, $c=1$. However, equation (\ref{last}) can be solved for any $q$ and $c$, giving origin to a large class $f(R)$ theories with polynomials roots. The exact solution also boils down to a quadrature of fractional functions, as one can see by substituting $R=e^\rho$ and $f=e^{\int h d\rho}$ in (\ref{last}), leading to the separable equation
 \begin{equation}
 h' = c^{-1}h^{1-n} - h^2 + h,
 \end{equation}
 where the tilde denotes differentiation with respect to $\rho$. Notice that for all $f(R)$ theories such that (\ref{last}), we have no problem with the invertibility of (\ref{u2u3}). However, this is a real 
 issue if the expression (\ref{u2u3}), as a function of $R$, fails to be monotonic. In this case, we would have  distinct function $\gamma(u_2,u_3)$ and, consequently, different equations of motion according to the
 value of $R$, which would complicate considerably the dynamics of the system.

\acknowledgements
The authors   thank  S. Pal and  J.D. Barrow for    enlightening discussions, and the Yukawa Institute for Theoretical Physics at Kyoto University, Kyoto, Japan, for the warm hospitality   during the long-term workshop YITP-T-17-02 ``Gravity and Cosmology 2018'', 
where this work was initiated.
A.S. is also grateful to FAPESP (grant 2013/09357-9) and CNPq for the financial support.
The work of K.B. was partially supported by the JSPS KAKENHI Grant Number JP
25800136 and Competitive Research Funds for Fukushima University Faculty
(18RI009).

\end{document}